\documentclass[%
 reprint,
superscriptaddress,
 amsmath,amssymb,
 aps,
 prl,
floatfix
]{revtex4-1}
\usepackage{pdfpages}
\usepackage{pgffor}
\usepackage{graphicx}% Include figure files
\usepackage{dcolumn}% Align table columns on decimal point
\usepackage{bm}% bold math
\usepackage{hyperref}
\usepackage{xr}
\usepackage{physics}
\usepackage{chemformula}
\usepackage{siunitx}
\usepackage{color,soul}
\usepackage{float}

\newcommand{\anni}{\hat{c}}
\newcommand{\crea}{\hat{c}^{\dagger}}
\newcommand{\annih}{\hat{b}}
\newcommand{\creah}{\hat{b}^{\dagger}}
\newcommand{\bk}{\mathbf{k}}
\newcommand{\cpkp}{c{'}\mathbf{k{'}}}

\newcommand{\bmk}{\mathbf{k}}
\newcommand{\bmkp}{\mathbf{k{'}}}
\newcommand{\bmq}{\mathbf{q}}
\renewcommand{\braket}[1]{\expval{#1}}

\makeatletter
\AtBeginDocument{\let\LS@rot\@undefined}
\makeatother

\begin{document}

\preprint{APS/123-QED}

    \title{First Principles Calculation of Ballistic Current from Electron-Hole Interaction}

    \author{Zhenbang Dai}
	\affiliation{Department of Chemistry, University of Pennsylvania, Philadelphia, Pennsylvania 19104--6323, USA}
    \author{Andrew M. Rappe}%
	\affiliation{Department of Chemistry, University of Pennsylvania, Philadelphia, Pennsylvania 19104--6323, USA}

\date{\today}

\begin{abstract}
The bulk photovoltaic effect (BPVE) has attracted an increasing interest due to its potential to overcome the efficiency limit of traditional photovoltaics, and much effort has been devoted to understanding its underlying physics. However, previous work has shown that theoretical models of the shift current and the phonon-assisted ballistic current in real materials do not fully account for the experimental BPVE photocurrent, and so other mechanisms should be  investigated in order to obtain a complete picture of BPVE. In this Letter, we demonstrate two approaches that enable the \textit{ab initio} calculation of the ballistic current originating from the electron-hole interaction in semiconductors. Using \ch{BaTiO3} and \ch{MoS2} as two examples, we show clearly that for them the asymmetric scattering from electron-hole interaction is less appreciable than that from electron-phonon interaction, indicating more scattering processes need to be included to further improve the BPVE theory. Moreover, our approaches build up a venue for predicting and designing materials with larger ballistic current due to electron-hole interactions.

\begin{description}
\item[Keywords]
BPVE, shift current, ballistic current, first principles, electron-hole interaction, exciton
\end{description}
\end{abstract}

\maketitle

\textit{Introduction.--}
The bulk photovoltaic effect (BPVE) describes the generation of dc photocurrent in a homogeneous material, in contrast to tradition photovoltaic effects where a heterojunction is usually needed.~\cite{Belinicher80p199}
Such phenomenon requires the breaking of inversion symmetry or time-reversal symmetry, but it is not restricted by the upper limit of efficiency imposed on traditional solar cells. The BPVE can provide a large open-circuit photovoltage, thus attracting an increasing interest in the past few years in the field of opto-electronics.~\cite{Belinicher88p29,Spanier16p611}
On the other hand, the underlying physics of the BPVE is still under debate.
A first-principles calculation suggested that the shift current~\cite{vonBaltz81p5590}, which describes the coordinate shift during the optical excitation process, is the dominant mechanism in BaTiO$_3$~\cite{Young12p116601}, but more recent first-principles studies showed that shift current only contributes a portion (perhaps still a majority) of the total BPVE current.~\cite{Fei20p045104} Therefore, other processes have to be considered as well in order to have a complete description of BPVE.

Another important mechanism for the BPVE has  been proposed and studied, the  \textit{ballistic current} (BC).~\cite{Belinicher80p199}
Ballistic current originates from the asymmetric carrier generation at $\mathbf{k}$ and $\mathbf{-k}$, which in turn will induce a net current. 
The asymmetric carrier generation can be attributed to coherent scatterings from multiple contributions, such as the electron-phonon interaction, electron-hole interaction, and defects.
%Various papers have suggested that one mechanism or another is the dominant explanation for the BPVE.
It was reported that the ballistic current due to electron-hole scattering largely rationalized the photocurrent near the band edge in GaAs.~\cite{Sturman1992}
Recently, we have shown that the ballistic current arising from the intrinsic electron-phonon interaction  can have comparable magnitude with the shift current,  enhancing the overall agreement of the theoretical and experimental BPVE spectra, though some discrepancy still persists.~\cite{Dai20pArXiv}
Thus, it is of great interest to explore other contributions to ballistic current and understand their importance toward the overall BPVE phenomenon.

In this Letter, we use first-principles calculations to investigate the asymmetric carrier generation in semiconductors from another intrinsic scattering process, the electron-hole scattering. We include Coulomb interactions between the electrons excited to the conduction band and the holes left in the valence band across the whole Brillouin zone, which goes beyond previous treatment where only band extremum states are considered.~\cite{Shelest79p1353}
This interaction is known to give rise to exciton states, which  greatly modify the optical properties of materials, and it can also lead to asymmetric scattering of carriers.~\cite{Onida01p601_1}
Due to the intimate relation to excitons, we will call the current from electron-hole interactions the \textit{exciton ballistic current} (ex-BC).
The current can be calculated within a Boltzmann transport model:
\begin{align}
\label{boltzmann}
    j^{\alpha\beta,\gamma}(\omega)
    &= 2e\tau_0 \sum_{cv\bmk} \Gamma_{cv,\bmk}^{\alpha\beta}(\omega) \bqty{ v_{c\bmk}^{e,\gamma} -
    v_{v\bmk}^{e,\gamma}}
\end{align}
where $\Gamma_{cv,\bmk}^{\alpha\beta}$ is the carrier generation rate for an electron-hole pair $(c,v)$ at $\mathbf{k}$, $e$ is the electron charge, $\tau_0$ is the momentum relaxation time, and $\mathbf{v}^e_{c\bmk}$ ($\mathbf{v}^e_{v\bmk}$) is the electron (hole) velocity obtained from band derivatives.
The main task is to calculate $\Gamma_{cv,\bmk}^{\alpha\beta}(\omega)$, and we  take two approaches to tackle this problem: a phenomenological model which has been widely used to describe Wannier excitons
\cite{Knox63,Combescot15excitons}, and a many-body approach that largely resembles the modern \textit{ab initio} way of computing exciton states.~\cite{Rohlfing98p2312,Onida01p601_1}
We implement these two approaches via first-principles theory and compare the results with shift current and phonon-assisted ballistic current (ph-BC). 
The first-principles results show clearly that for \ch{BaTiO3} and \ch{MoS2}, the ex-BC  only makes a minor contribution to the overall BPVE.

\begin{figure}
    \centering
    \includegraphics[width=80mm]{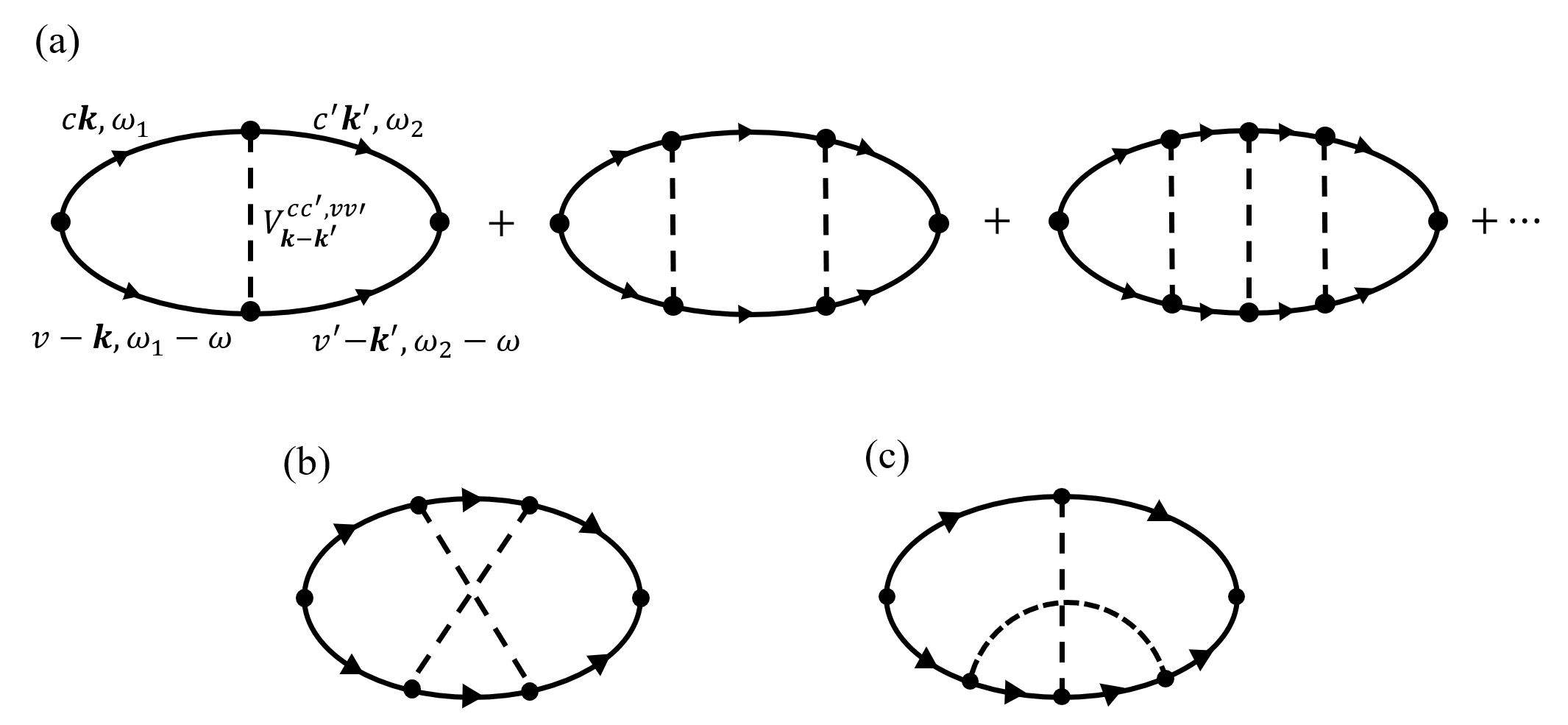}
    \caption{\label{fig:feynman}
    Feynman diagrams for electron-hole interaction. For a semiconductor, only ladder diagrams in (a) will contribute to asymmetric scattering.}
\end{figure}

\textit{Phenomenological Model.--}
We first derive an expression for ex-BC from a phenomenological treatment that is widely used when investigating Wannier excitons.
The model starts by assuming a single-particle Hamiltonian for the unperturbed state and taking the electron-hole interaction as a perturbation:
\begin{align}
\label{h0}
    H_0 = 
    \sum_{c\bk}\epsilon_{c\bk}\crea_{c\bk}\anni_{c\bk}
    +\sum_{v\bk}\epsilon_{v\bk}\creah_{v\bk}\annih_{v\bk}
\end{align}

\begin{align}
\label{int}
    V_{\rm int} = 
    -\sum_{cc'vv'}\sum_{\bk\bk'\bmq}
    {
    V_{\bmq}^{cc',vv'}
    {\crea_{c\bk+\bmq}}{\creah_{v-\bk'-\bmq}}
    {\annih_{v'-\bk'}}{\anni_{c'\bk}}
    },
\end{align}
where $\crea_{c\bk}$($\anni_{c\bk}$) and $\creah_{v\bk}$($\annih_{v\bk}$) are the creation (annihilation) operators for electrons and holes, respectively, $\epsilon_{n\bmk}$ is the eigenvalue of $H_0$, and $V_{\bmq}^{cc',vv'}$ is the screened Coulomb interaction in the basis of eigenstates of $H_0$.~\cite{Mahan13many} 
Note that $\creah_{v\bk} \equiv \anni_{v-\bk}$.
Then, in order to calculate the total carrier generation rate $\mathrm{\Gamma^{\alpha\beta}(\omega)}$, we need to evaluate the retarded momentum-momentum correlation function $\mathrm{\chi^{\alpha\beta}(\omega)}$ as prescribed in \cite{Dai20pArXiv}:
\begin{equation}
\label{carrier_generation}
    \Gamma^{\alpha\beta}(\omega) =
    -\frac{2}{\hbar}
    \Im\bqty{\chi^{\alpha\beta}(\omega)}
    \Big( {\frac{e}{m\omega}} \Big)^2 
    E_{\alpha}E_{\beta}.
\end{equation}

\begin{align}
\label{pp_correlation}
    &{\chi_{\rm T}^{\alpha\beta}}(i\omega_n) = -\frac{1}{\hbar} \sum_{\bmk cv}
    \braket{v\bmk|\hat{p}^{\alpha}|c\bmk}
    D^{\beta}_{cv,\bmk}(i\omega_n)
\end{align}
where
\begin{align}
\label{d_in_correlation}
    &D^{\beta}_{cv,\bmk}(i\omega_n) = 
    \sum_{\bmkp c'v'}
    \braket{c\bmk|\hat{p}^{\beta}|v\bmk} \nonumber \\
    &\times
    \int_{0}^{\hbar/k_{B}T} d{\tau} e^{i{\omega_n}{\tau}}
    \braket{\hat{T}_{\tau} {\annih_{v-\bmk}(\tau)} {\anni_{c\bmk}(\tau)} {\crea_{\cpkp}(0)} {\creah_{v'-\bmk'}(0)} }.
\end{align}
Here, momentum-momentum correlation functions $\mathrm{\chi^{\alpha\beta}(\omega)}$ (real-time, retarded) and $\mathrm{\chi_T}^{\alpha\beta}(i\omega_n)$ (imaginary-time) can be related via an analytical continuation:
${{\chi}^{\alpha\beta}}(\omega)={{\chi}_{\rm T}^{\alpha\beta}}({i\omega_n\xrightarrow[]{}\omega+i0^{+}})$ \cite{Mahan13many,Jishi13Feynman},
and from Eq.~\ref{pp_correlation} we can see that $\Gamma^{\alpha\beta}(\omega)$ can be decomposed as $\Gamma^{\alpha\beta}(\omega)=\sum_{cv,\bmk}
\Gamma_{cv,\bmk}^{\alpha\beta}(\omega)$.
The correlation function in Eq.~\ref{d_in_correlation} is evaluated under the full Hamiltonian $H=H_0+V_{\rm int}$, so to calculate it perturbatively, we carry out a diagrammatic approach by expressing each perturbation term as a Feynman diagram. Only ladder diagrams will contribute to the asymmetric scattering for insulators (\cite{Mahan67p882,Mahan13many}, and see SI).
However, due to the long-range character of the Coulomb interaction, we cannot  simply retain the lowest-order term. 
Thus, we need to sum up all orders of ladder diagrams as shown in Fig.~\ref{fig:feynman}.

After applying the Feynman rule for each diagram and going through a fair amount of algebra (see SI), the sum of the ladder diagrams can be written as:
\begin{align}
\label{vecD}
    \mathbf{D}_{cv,\bk}(\omega)
    &=
    \sum_{n=0}^{\infty}
    {\mathbf{D}^{(n)}_{cv,\bk}(\omega)} 
    \nonumber \\
    &=i
    \frac{\mathbf{\tilde{p}}_{cv,\bmk}(\omega)}
    {\omega+{\epsilon}_{v\bk}/\hbar-{\epsilon}_{c\bk}/\hbar+i0^{+}},
\end{align}
and
\begin{align}
\label{effective_p}
    &\mathbf{\tilde{p}}_{cv,\bmk}(\omega) 
    =\braket{c\bmk|\mathbf{\hat{p}}|v\bmk}
    \nonumber \\
    &+\sum_{\bmkp c'v'}{
    \frac{i}{\hbar}
    \frac{V_{\bk-\bk'}^{cc',vv'}}
    {\omega+{\epsilon}_{v'\bk'}/\hbar-{\epsilon}_{c'\bk'}/\hbar+i0^{+}}
    \mathbf{\tilde{p}}_{c'v',\bmk'}(\omega) 
    }.
\end{align}
This summation was first carried out in \cite{Mahan67p882}, where the effective mass approximation was assumed and the $\bmk$-dependence of the momentum matrix was neglected. 
The more general Eq.~\ref{effective_p} can be numerically solved for each frequency $\omega$ on a $\bmk$-grid, so together with Eq.~\ref{pp_correlation}, Eq.~\ref{vecD}, and Eq.~\ref{effective_p}, the carrier generation rate $\Gamma_{cv,\bmk}^{\alpha\beta}(\omega)$ can be calculated from first principles. 
We note here that a similar derivation for $\Gamma_{cv,\bmk}^{\alpha\beta}(\omega)$ has been carried out in \cite{Shelest79p1353,Sturman1992}, and fairly good agreement with experiments has been achieved. However, their focus is exclusively on the near-band-edge states, and only first-order perturbation is used when treating $\bmk$ points farther away from the band edge. Therefore, our derivation has the advantage of incorporating more states throughout the whole Brillouin zone and giving a current response for a broader frequency range.

\begin{figure*}
\includegraphics[width=170mm]{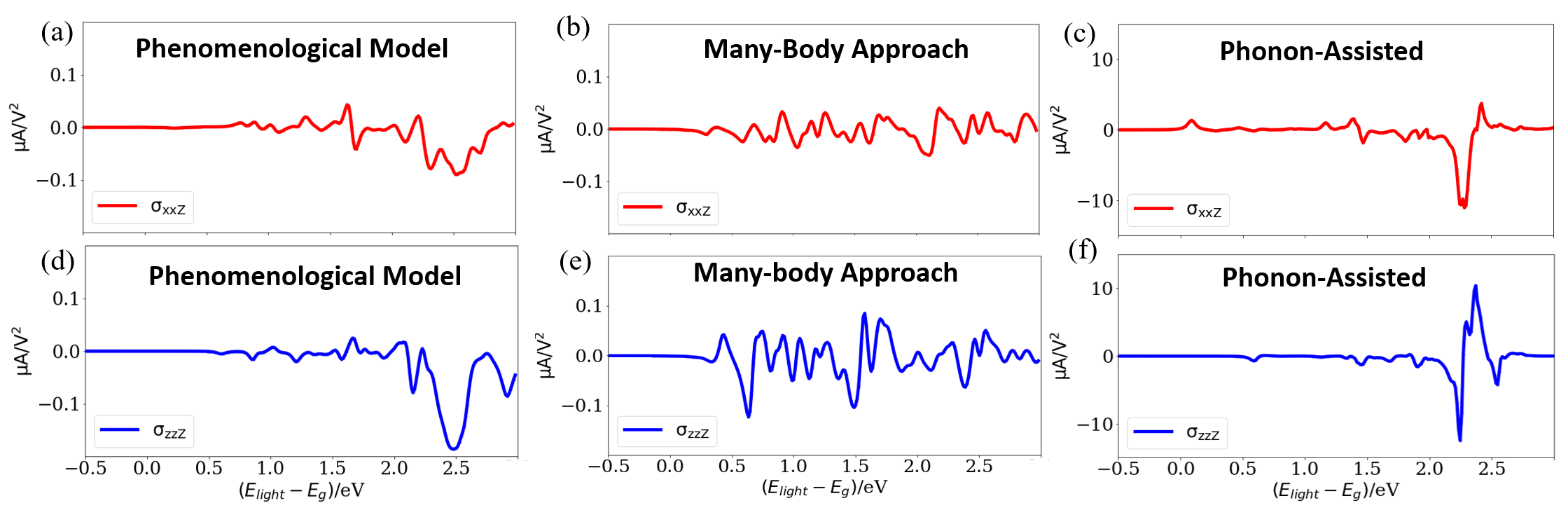}
\caption{\label{fig:BTO} First-principles results of ex-BC for tetragonal \ch{BaTiO3}. 
Note the scale difference in these figures.
(a)-(c) are the $xxZ$ component of the current response tensors, and (d)-(f) are the $zzZ$ component. Both the phenomenological model (a,d) and the many-body approach (b,e)  give results of similar magnitude, two orders smaller than the phonon-assisted ballistic current as shown in (c) and (f) (reproduced from~\cite{Dai20pArXiv}).
}
\end{figure*}

\textit{Many-body Approach.--}
Here we present another approach to calculate the carrier generation rate, in which the wave functions of exciton states $\ket{S}$ are computed explicitly from many-body theory.  According to Fermi's golden rule \cite{Bassani76p58}, the total carrier generation rate due to light absorption can be expressed as:
\begin{align}
\label{golden_rule}
    \Gamma^{\alpha\beta}(\omega)
    &=\frac{2\pi}{\hbar}
    \Big(\frac{e}{m\omega} \Big)^2
    E^{\alpha}E^{\beta} \nonumber \\
    &\times
    \sum_{S}
    \braket{0|\hat{p}^{\alpha}|S}
    \braket{S|\hat{p}^{\beta}|0}
    \delta (\Omega^S-\hbar\omega),
\end{align}
where $\Omega^{S}$ is the excitation energy and $\ket{0}$ is the ground state of the system.
The exciton states can be expressed by the linear combination of electron-hole pair states~\cite{Rohlfing98p2312}:
\begin{align}
\label{exciton_states}
    \ket{S}
    =\sum_{vc\bmk}
    {
    A_{vc\bmk}^S \ket{vc\bmk}
    }
    =\sum_{vc\bmk}
    {
    A_{vc\bmk}^S \crea_{c\bmk} \creah_{v-\bmk} \ket{0}
    },
\end{align}
and we call $A_{vc\bmk}$ the exciton wave functions. By substituting Eq.~(\ref{exciton_states}) into Eq.~(\ref{golden_rule}), we can rewrite the carrier generation rate as:
\begin{align}
\label{expand_golden}
    &\Gamma^{\alpha\beta}(\omega)
    =\frac{2\pi}{\hbar}
    \Big(\frac{e}{m\omega} \Big)^2
    E^{\alpha}E^{\beta} \nonumber \\
    &\times
    \sum_{vc\bmk} \sum_{S} 
    |A^{S}_{vc\bmk}|^2
    \braket{v\bk|\hat{p}^{\alpha}|c\bk}
    \braket{c\bk|\hat{p}^{\beta}|v\bk}
    \delta (\Omega^S-\hbar\omega),
\end{align}
from which we can see again that the overall generation rate can be split into $\sum_{cv,\bmk}
\Gamma_{cv,\bmk}^{\alpha\beta}(\omega)$.
Eq.~(\ref{expand_golden}) is to be contrasted with the carrier generation rate (transition rate) in a more well-known form that does not take electron-hole interaction into account:

\begin{align}
\label{old_transition}
    &\Gamma_{cv,\bmk}^{\alpha\beta,{\rm no}\_eh}(\omega)
    =\frac{2\pi}{\hbar}
    \Big(\frac{e}{m\omega} \Big)^2
    E^{\alpha}E^{\beta} \nonumber \\
    &\times
    \sum_{vc\bmk}
    \braket{v\bk|\hat{p}^{\alpha}|c\bk}
    \braket{c\bk|\hat{p}^{\beta}|v\bk}
    \delta (E_{c\bk}-E_{v\bk}-\hbar\omega).
\end{align}

Clearly, 
$\Gamma_{cv,\bmk}^{\alpha\beta,{\rm no}\_eh}(\omega)$ together with $\beta\alpha$ component
is symmetric under $\bmk \Leftrightarrow -\bmk$ and therefore gives no net current. 
Thus, it is the asymmetry lying in the exciton wave functions $A^{S}_{vc\bmk}$ that will make the carrier generation at $\bmk$ and $-\bmk$ unequal.

The exciton wave functions and excitation energies can be obtained by solving the \textit{Bethe-Salpeter Equation} (BSE):
\begin{align}
\label{bse}
(E_{c\bmk}-E_{v\bmk})A^{S}_{vc\bmk}+
\sum_{v'c'\bmk'}\braket{
vc\bmk|K^{eh}|v'c'\bmk'}
=\Omega^{S}A^{S}_{vc\bmk},
\end{align}
where the kernel $K^{eh}$ describes the electron-hole interaction, whose explicit form and technical details have been discussed extensively in \cite{deslippe12p1269}.
Once the BSE is solved, we can use $A^{S}_{vc\bmk}$ and $\Omega^{S}$ to calculate the carrier generation rate $\Gamma_{cv,\bmk}^{\alpha\beta}(\omega)$ via Eq.~(\ref{expand_golden}).

\textit{First-Principles Results.--}
The two approaches can be implemented numerically to calculate carrier generation rate $\Gamma_{cv,\bmk}^{\alpha\beta}(\omega)$, and together with Eq.~(\ref{boltzmann}), the exciton ballistic current for real materials can also be calculated via first principles.
To demonstrate this capability and in order for better comparison, we choose the prototypical BPVE material tetragonal \ch{BaTiO3} due to the availability of experimental BPVE data \cite{Koch75p847,Koch76p305} as well as first-principles results of shift current \cite{Young12p116601,Fei20p045104} and ph-BC.~\cite{Dai20pArXiv} In addition to \ch{BaTiO3}, we also applied the same calculation to the 2D material \ch{MoS2}, which has a strong exciton effect \cite{Wang20p11936} and is expected to possess a more obvious ex-BC. The structural data are obtained from \cite{Buttner92p764,Schankler21p1244}.

We perform density functional theory (DFT) calculations using the \textsc{Quantum Espresso} package.~\cite{Giannozzi09p395502, Giannozzi17p465901}
The generalized-gradient approximation exchange-correlation functional and norm-conserving pseudopotentials produced by the \textsc{Opium} package are used.~\cite{Perdew96p3865, Rappe90p1227, Ramer99p12471}
The convergence threshold for self-consistent calculations was \SI{e-8}{Ry/cell}. 
$GW$ and BSE calculations were performed using \textsc{BerkeleyGW} \cite{deslippe12p1269,Hybertsen86p5390,Rohlfing00p4927} to find the exciton wave functions. For illustrative purposes, all quantities are sampled on an \num{8x8x8}  k-grid for \ch{BaTiO3} and \num{16x16x1} for \ch{MoS2}  \cite{Monkhorst76p5188}, and denser grids can be used for finer spectral features. 
In the $GW$ calculations, the cut-off of the dielectric matrix was set as 10~Ry. For \ch{BaTiO3}, we include 20 valence and 200 conduction bands, whereas for \ch{MoS2}, 13 valence and 130 conduction bands are included. 
To construct the $K^{eh}$ in the BSE, 6 valence and 6 conduction bands are used in \ch{BaTiO3} and 9 valence and 9 conduction bands are used in \ch{MoS2}.
The relaxation times for \ch{BaTiO3} and \ch{MoS2} are chosen to be 2~fs and 100~fs, respectively.~\cite{Dai20pArXiv,Fei20p035440}

The results for \ch{BaTiO3} are shown in Fig.~\ref{fig:BTO}. It can be seen that both approaches will give ex-BC current of similar magnitude, though their spectral features  differ somewhat. The ex-BC magnitudes are 1--2 orders smaller than the shift current and ph-BC. From this, we can see that even though we have considered infinite order of e-h scattering perturbatively, their summation converges to a small value compared with ph-BC where only the lowest-order electron-phonon interaction is taken into account. Hence, for \ch{BaTiO3}, the electron-hole interaction only has a minor contribution to the asymmetric scattering, and one has to resort to other scattering mechanisms, such as  defect- or disorder-induced scattering in order to explain the discrepancy between experiments and simulations. 

\begin{figure}
    \centering
    \includegraphics[width=85mm]{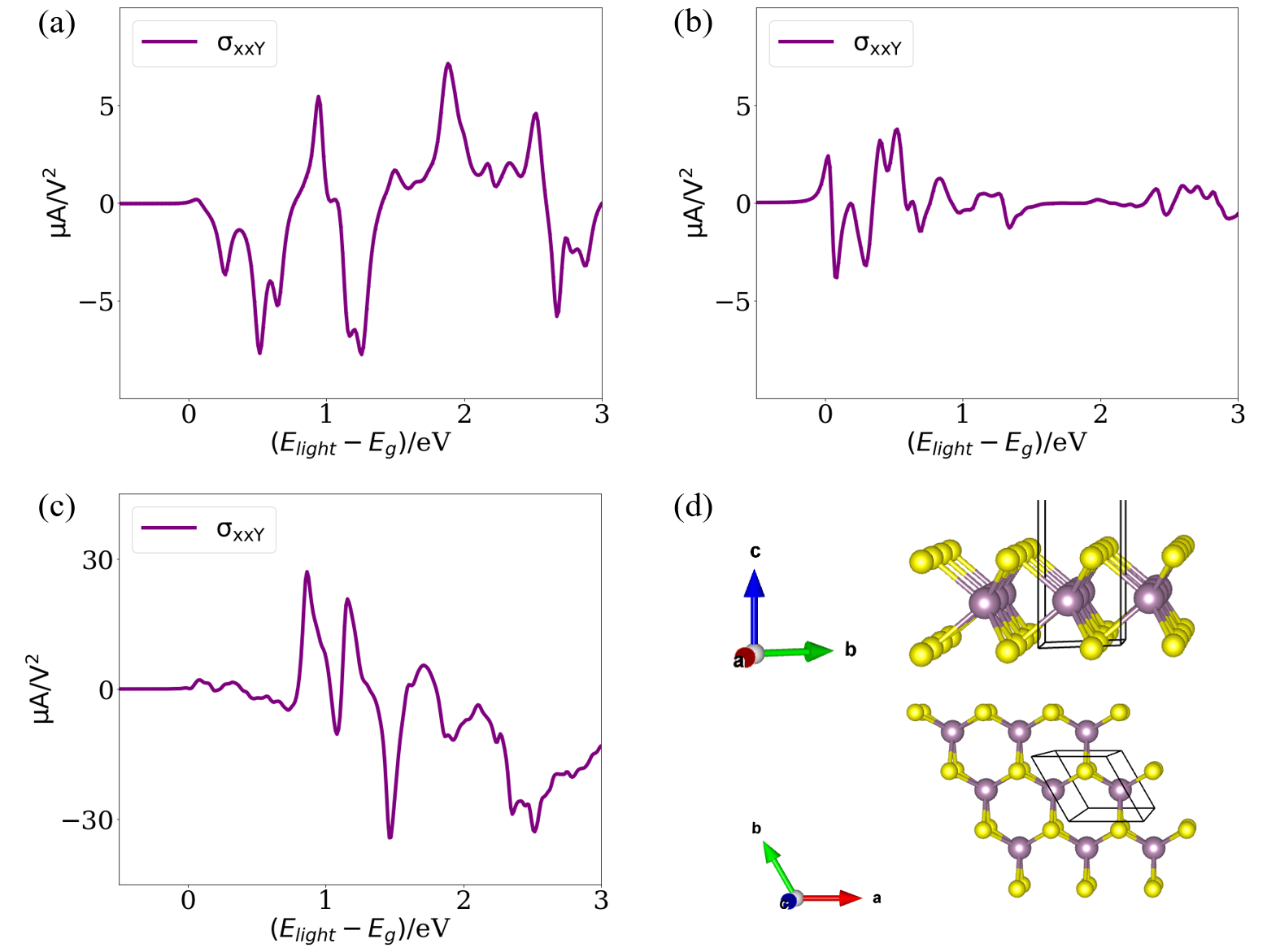}
    \caption{First-principles results for \ch{MoS2}. (a) Exciton ballistic current ex-BC from the phenomenological model. (b) ex-BC from the many-body approach. (c) Shift current as reproduced from \cite{Fei20p035440}. (d) Atomic structure of \ch{MoS2}.
    The \ch{MoS2} has a larger ex-BC due to the longer relaxation time[ref], but its asymmetric carrier generation rate is not necessarily large, even though strong exciton binding is expected for this material.
    }
    \label{fig:mos2}
\end{figure}

To see if electron-hole interaction has a greater impact on the asymmetric scattering for 2D materials which are expected to have strong exciton effects, we compared the ex-BC of \ch{MoS2} with its shift current, which has been reported to be large. In Fig.~\ref{fig:mos2}, it can be seen that even though the ex-BC is indeed larger than that of \ch{BaTiO3} due to a larger momentum relaxation time \cite{Fei20p035440}, its magnitude is still smaller than the shift current. Since the longer relaxation time, rather than a higher asymmetric carrier generation rate, causes the larger ex-BC in \ch{MoS2}, this shows  that  cancellations between various orders of ladder diagrams will make its overall contribution less appreciable, and any proper treatment of the Coulomb interaction should contain more than the lowest-order perturbation term.

\textit{Discussion and Conclusion.--}
The first-principles results for \ch{BaTiO3} and \ch{MoS2} show clearly that ex-BC is a minor contribution to the overall BPVE in these materials, so unlike the ph-BC which can greatly enhance the agreement between the theoretical BPVE spectra and experimental ones, the ex-BC is less significant, and perhaps it can be safely neglected when analyzing the constituents of experimental photocurrent for materials without a strong exciton effect. 
As a result, we would expect that defect or disorder scattering could make a stronger contribution to BPVE, not only because  it can induce asymmetric carrier generation, but also due to the coordinate shift that accompanies any scattering process, which resembles the shift current mechanism \cite{Sinitsyn06p075318,Xiao19p165422}. 

We would also like to comment on the relations between the two approaches we presented. 
The phenomenological model starts from a single-particle picture and treats the electron-hole interaction perturbatively, whereas the many-body approach is based on the full many-body picture (Green's function formalism and Hedin's equations) and can be reduced to a two-particle equation in the second iteration of solving Hedin's equations.~\cite{Onida01p601_1}
Therefore, the different nature at the very beginning makes it difficult to pinpoint exactly what causes the differences in the results, and this comparison deserves more rigorous investigation in the future.
A connection between the two approaches has been attempted by \cite{Sham66p708}, and they showed that when taking the effective-mass approximation, the two methods can be proved to be equivalent. 
When the full $\bmk$-dependence and multi-band nature are recovered, however, it becomes unclear how the two methods correspond to each other. 
Nevertheless, the similar magnitude of ex-BC from the two approaches confirm their overall validity, and together they illustrate the relative importance of electron-hole interaction to other asymmetric scattering mechanisms.
% is mostly used in physics community in order to understand more complicated phenomena, such as bi-excitons and exciton-polaritons as a result of its simplicity \cite{Combescot15excitons}, whereas the many-particle approach is more popular among the chemistry community where more accurate predictions are needed \cite{Onida01p601_1}.

To summarize, we employed two approaches to investigate the ballistic current from electron-hole scattering. Following these two methods, we performed first-principles calculations on \ch{BaTiO3} and \ch{MoS2} to demonstrate the influence of ex-BC on BPVE. Our numerical results show that electron-hole scattering has a relatively small contribution to BPVE for these two materials compared with shift current and phonon-assisted ballistic current, so more scattering mechanisms should be considered in order to better explain their experimental photocurrents.
On the other hand, the two approaches presented here enable the numerical evaluation of the importance of eh-BC for any semiconductor, and they open up the possibility for \textit{ab initio} prediction and design of materials with larger ex-BC.

{\em Acknowledgments.} This work was supported by the U.S. Department of Energy, Office of Science, Basic Energy Sciences, under Award \#
DE-FG02-07ER46431.
Computational support was provided by the National Energy Research Scientific Computing Center (NERSC), a U.S.
Department of Energy, Office of Science User Facility located at Lawrence Berkeley National Laboratory, operated under
Contract No. DE-AC02-05CH11231.
We acknowledge valuable discussions with Lingyuan Gao.

% \bibliography{BC}% Produces the bibliography via BibTeX.
\bibliography{rappecites}


\section*{Supplementary material}

# First Principles Calculation of Ballistic Current from Electron-Hole Interaction

Zhenbang Dai[1] and Andrew M. Rappe[1]

[1]*Department of Chemistry, University of Pennsylvania,*
*Philadelphia, Pennsylvania 19104–6323, USA*
(Dated: February 24, 2021)

## SUMMATION OF LADDER DIAGRAMS

In this section, we are going to show how the summation of ladder diagrams is carried out. As in [1], we will use the real-time Green's functions for electrons and holes because temperature will barely influence the occupation of a wide-gap semiconductor, which is what we are interested in, and the derivation can be generalized to finite temperature with imaginary-time Green's function. To get an idea what the summation looks like, we first apply the Feynman rule on the zeroth, first, and second order ladder diagrams:

$$D_{cv,\mathbf{k}}^{\beta(0)}(\omega) = \sum_{c'v'\mathbf{k}'} \delta_{c,c'} \delta_{v,v'} \delta_{\mathbf{k},\mathbf{k}'} \left\langle c'\mathbf{k}'|\hat{p}^\beta|v'\mathbf{k}'\right\rangle \int_{-\infty}^{+\infty} \frac{d\omega_1}{2\pi} G_c(\omega_1,\mathbf{k}) G_v(\omega_1-\omega,\mathbf{k})$$
$$= i \left\langle c\mathbf{k}|\hat{p}^\beta|v\mathbf{k}\right\rangle \frac{1}{\omega+\epsilon_{v\mathbf{k}}-\epsilon_{c\mathbf{k}}+i0^+} \tag{S1}$$

$$D_{cv,\mathbf{k}}^{\beta(1)}(\omega) = \sum_{c'v'\mathbf{k}'} \left\langle c'\mathbf{k}'|\hat{p}^\beta|v'\mathbf{k}'\right\rangle \frac{1}{\hbar} V_{\mathbf{k}-\mathbf{k}'}^{cc',vv'} \int_{-\infty}^{+\infty} \frac{d\omega_1}{2\pi} G_c(\omega_1,\mathbf{k}) G_v(\omega_1-\omega,\mathbf{k}) \int_{-\infty}^{+\infty} \frac{d\omega_2}{2\pi} G_{c'}(\omega_2,\mathbf{k}') G_{v'}(\omega_2-\omega,\mathbf{k}')$$
$$= \sum_{c'v'\mathbf{k}'} i^2 \left\langle c'\mathbf{k}'|\hat{p}^\beta|v'\mathbf{k}'\right\rangle \frac{1}{\hbar} V_{\mathbf{k}-\mathbf{k}'}^{cc',vv'} \frac{1}{\omega+\epsilon_{v\mathbf{k}}-\epsilon_{c\mathbf{k}}+i0^+} \frac{1}{\omega+\epsilon_{v'\mathbf{k}'}-\epsilon_{c'\mathbf{k}'}+i0^+} \tag{S2}$$

$$D_{cv,\mathbf{k}}^{\beta(2)}(\omega) = \sum_{c'v'\mathbf{k}'} \sum_{c''v''\mathbf{k}''} \left\langle c''\mathbf{k}''|\hat{p}^\beta|v''\mathbf{k}''\right\rangle \frac{1}{\hbar} V_{\mathbf{k}-\mathbf{k}'}^{cc',vv'} \frac{1}{\hbar} V_{\mathbf{k}'-\mathbf{k}''}^{c'c'',v'v''} \int_{-\infty}^{+\infty} \frac{d\omega_1}{2\pi} G_c(\omega_1,\mathbf{k}) G_v(\omega_1-\omega,\mathbf{k})$$
$$\times \int_{-\infty}^{+\infty} \frac{d\omega_2}{2\pi} G_{c'}(\omega_2,\mathbf{k}') G_{v'}(\omega_2-\omega,\mathbf{k}') \int_{-\infty}^{+\infty} \frac{d\omega_3}{2\pi} G_{c''}(\omega_3,\mathbf{k}'') G_{v''}(\omega_3-\omega,\mathbf{k}'')$$
$$= \sum_{c'v'\mathbf{k}'} \sum_{c''v''\mathbf{k}''} i^3 \left\langle c''\mathbf{k}''|\hat{p}^\beta|v''\mathbf{k}''\right\rangle \frac{1}{\hbar^2} V_{\mathbf{k}-\mathbf{k}'}^{cc',vv'} \frac{1}{\hbar} V_{\mathbf{k}'-\mathbf{k}''}^{c'c'',v'v''}$$
$$\times \frac{1}{\omega+\epsilon_{v\mathbf{k}}-\epsilon_{c\mathbf{k}}+i0^+} \frac{1}{\omega+\epsilon_{v'\mathbf{k}'}-\epsilon_{c'\mathbf{k}'}+i0^+} \frac{1}{\omega+\epsilon_{v''\mathbf{k}''}-\epsilon_{c''\mathbf{k}''}+i0^+} \tag{S3}$$

Note, unlike the main text, to simplify notation, $\epsilon_{c\mathbf{k}}$ and $\epsilon_{v\mathbf{k}}$ are in the unit of frequency in this Supplementary Material. Thus, we can see that each ladder diagram has a simple structure so that n-th order ladder diagram can be written as:

$$D_{cv,\mathbf{k}}^{\beta(n)}(\omega) = i \frac{1}{\omega+\epsilon_{v\mathbf{k}}-\epsilon_{c\mathbf{k}}+i0^+} \sum_{c_1 v_1 \mathbf{k}_1} ... \sum_{c_n v_n \mathbf{k}_n} \left(\frac{i}{\hbar}\right)^n \left\langle c_n \mathbf{k}_n|\hat{p}^\beta|v_n \mathbf{k}_n\right\rangle V_{\mathbf{k}-\mathbf{k}_1}^{cc_1,vv_1} V_{\mathbf{k}_1-\mathbf{k}_2}^{c_1 c_2,v_1 v_2} ... V_{\mathbf{k}_{n-1}-\mathbf{k}_n}^{c_{n-1}c_n,v_{n-1}v_n}$$
$$\times \prod_{j=1}^n \frac{1}{\omega+\epsilon_{v_j \mathbf{k}_j}-\epsilon_{c_j \mathbf{k}_j}+i0^+}$$
$$= i \frac{1}{\omega+\epsilon_{v\mathbf{k}}-\epsilon_{c\mathbf{k}}+i0^+} \mathscr{P}_{cv}^{\beta(n)}(\mathbf{k},\omega) \tag{S4}$$

where

$$\mathscr{P}_{cv}^{\beta(n)}(\mathbf{k},\omega) = \sum_{c_1 v_1 \mathbf{k}_1} ... \sum_{c_n v_n \mathbf{k}_n} \left(\frac{i}{\hbar}\right)^n \left\langle c_n \mathbf{k}_n|\hat{p}^\beta|v_n \mathbf{k}_n\right\rangle V_{\mathbf{k}-\mathbf{k}_1}^{cc_1,vv_1} V_{\mathbf{k}_1-\mathbf{k}_2}^{c_1 c_2,v_1 v_2} ... V_{\mathbf{k}_{n-1}-\mathbf{k}_n}^{c_{n-1}c_n,v_{n-1}v_n}$$
$$\times \prod_{j=1}^n \frac{1}{\omega+\epsilon_{v_j \mathbf{k}_j}-\epsilon_{c_j \mathbf{k}_j}+i0^+}. \tag{S5}$$

Therefore, the sum of all orders of ladder diagrams will be:

$$D_{cv,\mathbf{k}}^{\beta}(\omega) = \sum_{n=0}^{\infty} D_{cv,\mathbf{k}}^{\beta(n)}(\omega) = i \frac{1}{\omega+\epsilon_{v\mathbf{k}}-\epsilon_{c\mathbf{k}}+i0^+} \sum_{n=0}^{\infty} \mathscr{P}_{cv}^{\beta(n)}(\mathbf{k},\omega) \tag{S6}$$

Since the summation is up to infinite order, we can denote $\sum_{n=0}^{\infty} \mathscr{P}_{cv}^{\beta(n)}(\mathbf{k},\omega)$ as $\tilde{\mathbf{p}}_{cv,\mathbf{k}}(\omega)$, and a close inspection will reveal:

$$\tilde{\mathbf{p}}_{cv,\mathbf{k}}(\omega) = \left\langle c\mathbf{k}|\hat{p}|v\mathbf{k}\right\rangle + \sum_{c'v'\mathbf{k}'} \frac{i}{\hbar} \frac{V_{\mathbf{k}-\mathbf{k}'}^{cc',vv'}}{\omega+\epsilon_{v'\mathbf{k}'}-\epsilon_{c'\mathbf{k}'}+i0^+} \tilde{\mathbf{p}}_{c'v',\mathbf{k}'}(\omega). \tag{S7}$$

To verify this result, let's calculate the contribution of the zeroth and first order ladder diagram by Eq. S7:

$$\tilde{\mathbf{p}}_{cv,\mathbf{k}}(\omega) = \langle c\mathbf{k}|\hat{\mathbf{p}}|v\mathbf{k}\rangle + \sum_{c'v'\mathbf{k}'} \frac{i}{\hbar} \frac{V_{\mathbf{k}-\mathbf{k}'}^{cc',vv'}}{\omega + \epsilon_{v'\mathbf{k}'} - \epsilon_{c'\mathbf{k}'} + i0^+} \tilde{\mathbf{p}}_{c'v',\mathbf{k}'}(\omega)$$

$$= \langle c\mathbf{k}|\hat{\mathbf{p}}|v\mathbf{k}\rangle + \sum_{c'v'\mathbf{k}'} \frac{i}{\hbar} \langle c'\mathbf{k}'|\hat{\mathbf{p}}|v'\mathbf{k}'\rangle \frac{V_{\mathbf{k}-\mathbf{k}'}^{cc',vv'}}{\omega + \epsilon_{v'\mathbf{k}'} - \epsilon_{c'\mathbf{k}'} + i0^+}$$

$$+ \sum_{c'v'\mathbf{k}'} \sum_{c''v''\mathbf{k}''} \frac{i}{\hbar} \frac{V_{\mathbf{k}-\mathbf{k}'}^{cc',vv'}}{\omega + \epsilon_{v'\mathbf{k}'} - \epsilon_{c'\mathbf{k}'} + i0^+} \frac{V_{\mathbf{k}'-\mathbf{k}''}^{c'c'',v'v''}}{\omega + \epsilon_{v''\mathbf{k}''} - \epsilon_{c''\mathbf{k}''} + i0^+} \tilde{\mathbf{p}}_{c''v'',\mathbf{k}''}(\omega)$$

$$= \tilde{\mathbf{p}}_{cv,\mathbf{k}}^{(0)}(\omega) + \tilde{\mathbf{p}}_{cv,\mathbf{k}}^{(1)}(\omega) + ...$$

$$D_{cv,\mathbf{k}}^{\beta(0)}(\omega) + D_{cv,\mathbf{k}}^{\beta(1)}(\omega) = i\frac{1}{\omega + \epsilon_{v\mathbf{k}} - \epsilon_{c\mathbf{k}} + i0^+} (\tilde{p}_{cv,\mathbf{k}}^{\beta(0)}(\omega) + \tilde{p}_{cv,\mathbf{k}}^{\beta(1)}(\omega))$$

$$= i\frac{1}{\omega + \epsilon_{v\mathbf{k}} - \epsilon_{c\mathbf{k}} + i0^+} \left[ \langle c\mathbf{k}|\hat{p}^\beta|v\mathbf{k}\rangle + \sum_{c'v'\mathbf{k}'} \frac{i}{\hbar} \langle c'\mathbf{k}'|\hat{p}^\beta|v'\mathbf{k}'\rangle \frac{V_{\mathbf{k}'-\mathbf{k}''}^{c'c'',v'v''}}{\omega + \epsilon_{v''\mathbf{k}''} - \epsilon_{c''\mathbf{k}''} + i0^+} \right],$$
$$\text{(S8)}$$

which is the same as (Eq. S1)+(Eq. S2).

## NON-LADDER DIAGRAMS

Here, we present a few examples of non-ladder diagrams and show why they will not contribute to the asymmetric scattering for a semiconductor. For Fig. 1(b), applying Feynman rule will lead to:

$$I_b = \mathcal{N}_b \int_{-\infty}^{+\infty} d\omega_2 \frac{1}{\omega_2 - \epsilon_1 + i0^+} \frac{1}{\omega_3 - \omega + \omega_1 - \omega_2 - \epsilon_2 - i0^+}, \tag{S9}$$

$\mathcal{N}_b$ containing all other terms. The integral in Eq. S9 will be zero since all the poles for $\omega_2$ are lying in the lower-half plane and thus the upper-half plane will be analytic. As a result, choosing a contour that goes along x-axis from $-\infty$ to $+\infty$ and then goes back to $-\infty$ along a semicircle in the upper plane, the contour integral will be zero because of the analyticity, and due to the residue theorem, the integral in Eq. S9 will also be zero.

Similarly, for Fig. 1(c), applying the Feynman rule will give:

$$I_c = \mathcal{N}_c \int_{-\infty}^{+\infty} d\omega_3 \frac{1}{\omega_3 - \epsilon_1 - i0^+} \frac{1}{\omega_3 + \omega_1 - \omega_2 - \epsilon_2 - i0^+}, \tag{S10}$$

and similar argument can be made to prove that $I_c = 0$ as well. It can be noted that most non-ladder diagrams will make the poles appear only on one-half plane, which in turn will render a vanishing contribution for semiconductors.

However, there exists a class of non-ladder diagrams that will have finite contribution [2], which is in the form of Fig. S1. However, it is immediately realized that the summing over the motifs within the "bubble" will give rise to the RPA dielectric function [3]. Therefore, they can be absorbed in a Coulomb interaction line that is screened by the dielectric function, which has already been taken into account when constructing the phenomenological model where the Coulomb interaction in $\hat{V}_{int}$ is a screened one. We used the macroscopic dielectric constant to approximate this dielectric function, and it can be obtained from first principles calculations [4].

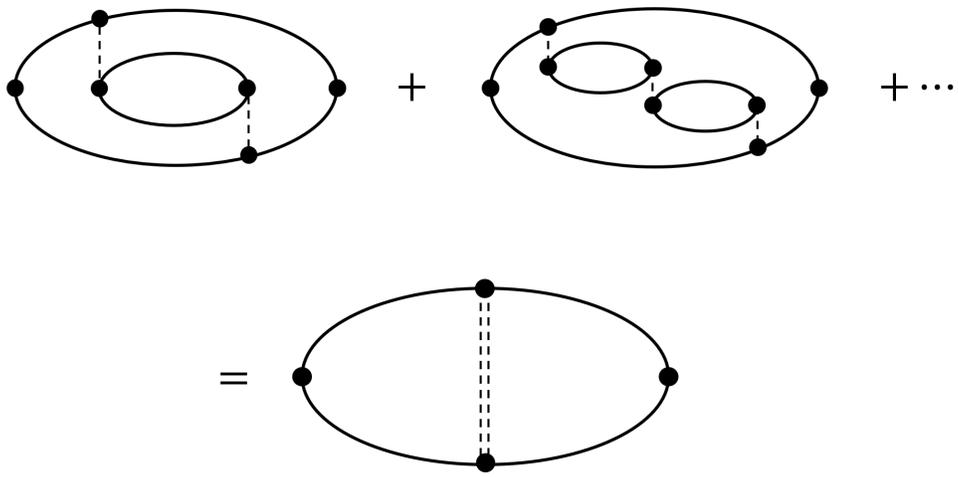

FIG. S1. A class of non-ladder diagrams that will give nonzero contribution. The summation of them will simply give a screened Coulomb interaction.


\end{document}